# Continuous Reinforcement Learning-based Dynamic Difficulty Adjustment in a Visual Working Memory Game

Masoud Rahim
mr.rahimi39@ut.ac.ir

Hadi Moradi
moradih@ut.ac.ir

Abdol-hossein Vahabie
h.vahabie@ut.ac.ir

Hamed Kebriaei
kebriaei@ut.ac.ir

Department of Electrical and Computer Engineering, University of Tehran, Tehran, Iran

*Abstract*—Dynamic Difficulty Adjustment (DDA) is a viable approach to enhance a player's experience in video games. Recently, Reinforcement Learning (RL) methods have been employed for DDA in non-competitive games; nevertheless, they rely solely on discrete state-action space with a small search space. In this paper, we propose a continuous RL-based DDA methodology for a visual working memory (VWM) game to handle the complex search space for the difficulty of memorization. The proposed RL-based DDA tailors game difficulty based on the player's score and game difficulty in the last trial. We defined a continuous metric for the difficulty of memorization. Then, we consider the task difficulty and the vector of difficulty-score as the RL's action and state, respectively. We evaluated the proposed method through a within-subject experiment involving 52 subjects. The proposed approach was compared with two rule-based difficulty adjustment methods in terms of player's score and game experience measured by a questionnaire. The proposed RL-based approach resulted in a significantly better game experience in terms of competence, tension, and negative and positive affect. Players also achieved higher scores and win rates. Furthermore, the proposed RL-based DDA led to a significantly less decline in the score in a 20-trial session.

## I. Introduction

The notion of difficulty adaptation, which is the act of tailoring the difficulty of a given task to match the ability of the person performing it, has been extensively discussed in a range of fields, including e-learning [1], rehabilitation [2], and video games [3]. Zohaib [4] mentioned that "DDA is a technique of automatic real-time adjustment of scenarios, parameters, and behaviors in video games, which follows the player's skill and keeps them from boredom (when the game is too easy) or frustration (when the game is too difficult)."

Based on Csikszentmihalyi's flow theory [5], a balance between the challenge of a task and a performer's ability is necessary to avoid boredom and frustration, which respectively result from too easy and too difficult tasks. Paraschos's and Koulouriotis's review paper [6] asserted that difficulty adaptation and personalization would significantly increase players' engagement and satisfaction. Also, many empirical studies, like [7] and [8], showed the efficacy of DDA in improving user experience, engagement, and enjoyment. Moreover, Vygotsky's [9] theory of the zone of proximal development (ZPD) supports the essence of difficulty adaptation. ZPD suggests that students can learn more effectively when their assignments are slightly more difficult than their current ability level. For example, students who played an adaptive educational video game achieved significantly better learning outcomes than those who played a non-adaptive game [10]. Several studies, like [11] and [12], empirically reported significant working memory capacity improvements in adaptive training compared to non-adaptive training. Additionally, another theory by Lövdén et al. [13] mentioned that "adult cognitive plasticity is driven by a prolonged mismatch between functional organismic supplies and environmental demands." In other words, a mismatch between ability and difficulty can lead to cognitive plasticity. Therefore, DDA is crucial to enhance the player experience and improve learning and cognitive training outcomes.

RL-based DDA approaches differ depending on whether the game is competitive or non-competitive. In competitive games, such as combat games or shooting games, players oppose an opponent in an activity, and the opponent's expertise determines the game difficulty. On the other hand, non-competitive games have no opponent, and players must accomplish specific tasks to win or receive a score, such as puzzles and working memory games.

Almost in all competitive games, RL-based DDA was accomplished by adjusting the reward or actions of the RL-based opponent, resulting in changes in the expertise of non-human opponents. For instance, Sithungu and Ehlers [14] used Q-learning to adjust the game difficulty by changing the decision-making level of the non-human player in each game to minimize the goal difference in a football game. Baláž and Tarábek [15] introduced a new action selection mechanism that allowed a super-human player agent to adapt to a weaker player by aiming to end the game in a draw in AlphaZero. In a turn-based battle video game, Pagalyte et al. [16] used SARSA to implement an RL agent whose reward was swapped automatically throughout the battle to encourage or discourage the non-human player agent from winning in order to maintain a good balance between challenges and skill levels. Noblega et al. [17] develop an RL-based non-player character that could play the game and adapt itself to other non-human players. They maintained the game balance in a fighting game by using a reward function based on a game-balancing constant.

However, there is no opponent in non-competitive games. Thus, in such games, an RL-based DDA is accomplished by changing the game elements and features. For instance, Huber et al. [18] utilized Deep RL techniques to modify a design of a maze design and determine the exercise rooms that should be incorporated into the maze. Reis et al. [19] applied DDA in a 2D grid maze that contains one movable wall. An RL agent, as the master agent in the game, tried to move the movable wall in order to achieve a balance for player agents with different proficiency. Spatharioti et al. [20] used Q-learning to create a series of levels with varying degrees of difficulty in an image-matching game. The RL-based DDA selected the difficulty of the next level from easy, medium, and hard. They showed that by using an RL-based DDA, the 150 subjects, who played the game, could collect more tiles, make more moves, and play the game longer than the game with no DDA. Zhang and Goh [21] utilized bootstrapped policy gradient (in order to improve sample efficiency) and multi-armed bandit RL. Moreover, offline clustering and online classification were proposed to capture users' differences in a VWM game. This game is very similar to the game discussed in this paper. The RL action was to select a memory task from the task bank that contained 100 tasks. They compared their RL-based DDA approach with a rule-based and a random approach, demonstrating its effectiveness in providing better adaptation for slow players in terms of players' performance.

To the best of our knowledge, previous studies on RL-based DDA in non-competitive games have relied solely on discrete state-action space with a small search space. In this study, to fill the mentioned gap, we propose an RL-based DDA with continuous state and action in a VWM game to handle the complex search space for the difficulty of memorization. The RL-based DDA tailors game difficulty for each player based on their score and game difficulty in the last trial. To this end, a continuous metric for the difficulty of memorization is defined by calculating the linear combination of three features of the memory task. Then, we consider the task difficulty and the vector of difficulty-score as the RL's action and state, respectively. The reward is based on the player's score. Additionally, we employ Proximal Policy Optimization (PPO) [22] algorithm to train the RL system. The training is performed through simulations of human players, followed by a fine-tuning phase using human subjects. We conducted an experiment on 52 healthy subjects and compared the proposed RL-based DDA with two other rule-based difficulty adjustment methods in terms of subjects' performance and subjective game experiences using a questionnaire. Our approach led to a significantly better game experience in terms of competence, tension, and negative and positive affect. The main contributions of this paper are as follows:

- We proposed a continuous metric for the difficulty of memorization in a VWM game whose search space for the difficulty is complex. Using this metric for DDA, players experienced a significantly less decline in the score as the game became more challenging in a 20-trial training session.
- We formulated the DDA issue in the VWM game using PPO as a continuous RL problem. Then, the training was performed through simulations of human players, followed by a fine-tuning phase using human subjects.
- The proposed RL-based DDA led to an enhanced game experience when compared to two other rule-based adaptation methods. It was assessed in a within-subject experiment involving 52 subjects who answered a questionnaire.

II. VISUAL WORKING MEMORY

Working memory (WM) is the brain's ability to memorize and recall information temporarily [23]. WM is a crucial element in many high-order cognitive processes [24]. Our VWM task has two stages: memorization and recall. In the memorization stage, a 6*6 hexagonal grid is displayed for two seconds, with certain hexagons simultaneously highlighted in yellow (known as "targets") while the rest are white. After 2 seconds, all the hexagons turn white, and the player must recall and click on the exact locations of the targets (recall stage). Correct clicks turn green and incorrect ones turn red. The player's score is the ratio of correct clicks to the total number of targets in that task, ranging from 0 to 1. A score of 1 represents a win, whereas other scores represent a loss (Fig. 1).

A. Difficulty of the Visual Working Memory Game

We should first determine task difficulty to adjust the difficulty to the player's ability. Previous studies usually considered the number of items, that a subject should memorize and recall, as the difficulty indicator of the memory task. In some WM tasks, like the memory span task, it is reasonable to consider the number of items as the difficulty indicator. However, in some WM tasks like VWM, the difficulty is not only influenced by the number of items, but other factors should be considered [21]. Specifically, the visual load of items, such as the orientation and the distribution of target cells as well as the number of items, is important [25], [26], [27]. Xu and Chun [28] showed that it is easier to memorize targets that can be spatially grouped than cases that cannot be grouped in chunks in VWM tasks. Treating each unique memory task image as one level could potentially solve the challenge of defining an exact difficulty metric in VWM games. However, the exponential growth of the distinct memory tasks, based on the number of cells, would empirically make it hard or impossible when the number of cells goes over. To solve this problem, in [25], the authors saved 100 distinct memory tasks in a database and defined each unique task as one level. In other words, they sampled 100 distinct tasks from the large space of memory tasks. However, this approach limits the variety of memory tasks and users may receive repetitive memory tasks, especially when players tend to perform continuous training.

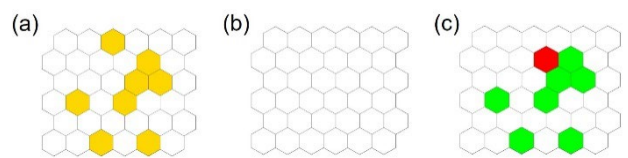

Fig. 1. Game interface. (a) Memorization stage in which eight targets are simultaneously shown to the subject for two seconds. (b) Recall stage in which player should click on the exaxt location of targets (c) Feedback for players in which correct and incorrect answers are displayed instantly (one incorrect and seven correct answers).

To address the shortage of a proper difficulty metric for the VWM game, we defined a continuous difficulty metric for the VWM game. This was achieved by extracting the top three features that serve as the best indicators of the difficulty: the number of targets ($n_t$), the number of connected components ($n_c$), and the distribution ($d$). These features are identified based on our experience and interviews with subjects. The continuous difficulty metric is defined as a normalized linear combination of these three features (Eq. 1).

$$difficulty = \alpha_1 n_t + \alpha_2 n_c + \alpha_3 d \quad (1)$$

The distribution feature is defined to quantify the sparsity of the target hexagons. The target cells in our VWM task are considered as the items in a matrix $D$ of size $n_t \times n_t$. Distance $d_{ij}$ represents the number of hexagons present in the shortest path between the $i^{th}$ and $j^{th}$ items in the memory task. It is a zero-diagonal matrix since $d_{ii}$s are zero. The distribution can be calculated by summing and normalizing the upper diagonal elements of $D$ (Fig. 2).

### III. RL-BASED DDA METHOD

We proposed a continuous RL-based DDA in our VWM game (Fig. 3). The RL-based DDA tailors the difficulty of the memory tasks based on a player's performance in the last trials. The objective is to keep players within a specific range of difficulty that aligns with their competence level. The RL action determines the difficulty of the next memory task. We considered the vector of the task difficulty and the player's score as the state of the RL. We defined three reward functions, which are formulated in equations (2), (3), and (4) respectively. Reward functions $R_1$ and $R_2$ were based on the eighty five percent rule [29] with adjustments to suit the nature of our VWM game. Additionally, $R_3$ was defined heuristically to maximize the score and task difficulty. Then, the RL system was trained with each reward function, followed by empirical testing through gameplay. $R_1$ significantly outperformed the other two reward functions in terms of the average reward and testers' feedback. The RL system was initialized assuming medium difficulty for our VWM task.

The other critical sub-element of the RL system is the environment which, in our case, is the variety of human players in the VWM game. Due to expensive exploration in the real environment, the RL system was trained in the environment simulation, followed by a fine-tuning phase using human subjects. We utilized hand-crafted player agents with various proficiency as environment simulation. The discount factor was equal to 0.95. The PPO algorithm was chosen after comparing it with other RL algorithms like Actor-critic, Deep Deterministic Policy Gradient (DDPG), Soft Actor-Critic, Twin Delayed DDPG, and Hindsight Experience Replay in terms of average reward in the simulation.

$$R_1(score) = \begin{cases} 0 & 0.9 < score \leq 1 \\ +1 & 0.7 \leq score \leq 0.9 \\ -1 & 0 < score < 0.7 \end{cases} \quad (2)$$

$$R_2(score) = \begin{cases} -0.5 & 0.9 < score \leq 1 \\ +1 & 0.7 \leq score \leq 0.9 \\ -1 & 0.4 \leq score < 0.7 \\ -2 & 0 \leq score < 0.4 \end{cases} \quad (3)$$

$$R_3(score, difficulty) = 0.35 \, score \times difficulty \quad (4)$$

When a continuous difficulty metric is defined, it may be necessary to create a lookup table that links the task difficulty to the corresponding game state. In our VWM game, it is possible to extract the features and calculate the difficulty of any memory task, but it is impossible to construct a memory task with a given difficulty value since it is not reversible. In our case, all 8,348,891,641 possible memory tasks, with $n_t$ ranging from 4 to 14, and their corresponding difficulty value are stored in the Difficulty Database (DDB). Once the RL agent determines the task difficulty, the memory task is shown by looking up the DDB.

### IV. EXPERIMENT

A within-subject experiment was conducted to evaluate the proposed RL-based DDA and the continuous difficulty metric. We compared the game experience and scores of 52 healthy subjects in three difficulty adjustment methods: RL-based, rule-based 1, and rule-based 2. Participants' game experience, including immersion, flow, competence, positive and negative affect, tension, and challenge, was assessed by the in-game module of the Game Experience Questionnaire (GEQ) [30].

Both rule-based difficulty adjustment methods have the same policy for difficulty adjustment, determined by hand-crafted rules. The only difference between these rule-based methods lies in their indicator of task difficulty. The first rule-based method determines difficulty based solely on the number of targets, such that all tasks with the same number of targets are considered equally challenging. For example, if the number of targets is eight, eight hexagon cells are randomly selected to become target cells. In the second rule-based method, the difficulty indicator is the continuous metric that we defined in this study and was explained in the Visual Working Memory section. This continuous difficulty metric was also used in the RL-based approach. In the first (second) rule-based method, the difficulty is increased by 1 (0.1) if the player scores above 90%, decreased by 1 (0.1) if the player scores below 70%, and is kept constant if the score falls between 70% and 90%. Both rule-based methods were initialized by the medium difficulty according to their metric.

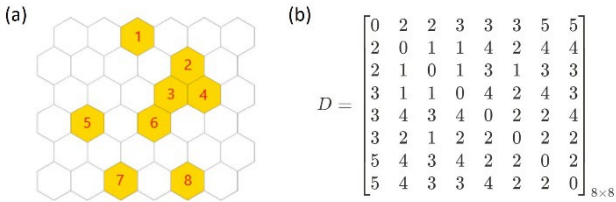

Fig. 2. (a) A memory task in which targets are numbered row-wise from left to right. (b) The corresponding matrix D of the memory task; $d_{18} = 5$ means there are five hexagons in the shortest path from cell number 1 to cell number 8 in the left image.

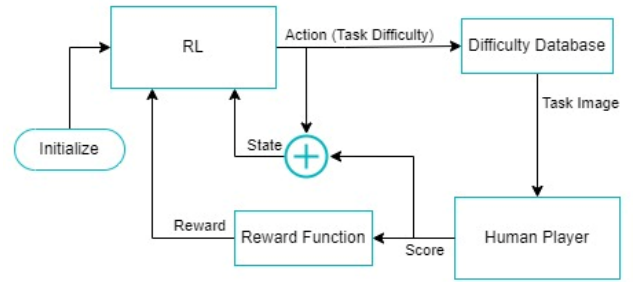

Fig. 3. The proposed RL-based DDA approach.

The game was introduced to each subject individually to ensure that participants were familiar with the game and the experiment[1]. Then, participants played a few training trials before starting the main experiment. The experiment consisted of playing our VWM game with the three difficulty adjustment methods for 20 trials per method. After completing each set of the 20 trials, participants were asked to fill out the GEQ, which provided valuable insights into their subjective experience of the game. Each participant played our VWM game with one difficulty adjustment method, answered the questionnaire, moved on to the next method, and repeated the same process until all three methods had been played (Fig. 4). To avoid any potential order effects on the results, we randomized the order of the difficulty adjustment methods for each participant.

Participants (30 men and 22 women aged between 19 and 32) were recruited from the School of Electrical and Computer Engineering, at the University of Tehran. The Research Ethics Committees of the school of Psychology and Education approved the study with an Ethics ID[2] of IR.UT.PSYEDU.REC.1401.107. The experiment was conducted in a controlled environment using a 15-inch laptop. Each participant was alone in the room during the experiment without a cell phone.

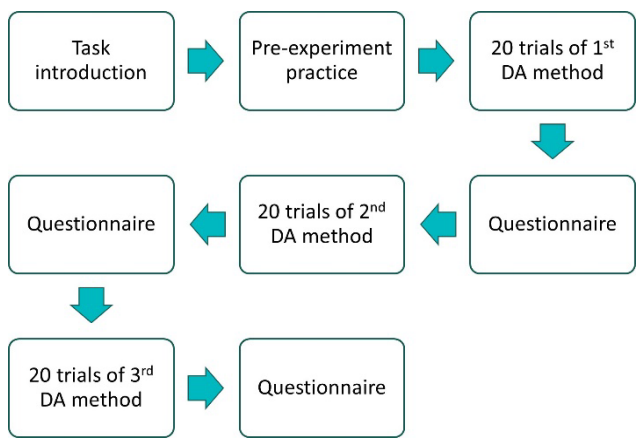

Fig. 4. The procedure of the experiment for each subject. DA is abbreviation for difficulty adjustment.

## V. Results

### A. Evaluation of DDA methods

#### 1) Analysis of the Questionnaire Data

The proposed RL-based DDA was compared with two the rule-based adaptation methods in terms of game experience. The results indicated a significantly better game experience in terms of competence, tension, and positive and negative affect for players when playing the RL-based DDA. Also, rates of immersion and flow were better than the other two rule-based methods, but not significantly (Fig. 5). The differences between the three groups were assessed using the Friedman test (Table I).

#### 2) Analysis of Gameplay Data

All 52 subjects participated the three difficulty adjustment methods for 20 trials per method. The average score was calculated for each subject in each method ($Avg_{s, m}$). It resulted in three vectors, each consisting of 52 elements. Each vector represents the average scores of the 52 subjects per method. The analysis of $Avg_{s, m}$ suggested that the proposed RL-based DDA resulted in significantly higher win rates and game scores for the participants than rule-based methods (Fig. 6). Each of the two methods was compared using paired t-tests (Table II).

The average score was calculated for each trial in each method ($Avg_{t, m}$). It resulted in 3 vectors, each consisting of 20 elements. Each vector represents the average scores of 20 trials per method (Fig. 7). According to Fig. 7, all three adjustment methods are able to keep the players in the desired score range while increasing the game difficulty. However, the RL-based method adjusted the difficulty so that subjets scored higher as the game difficulty increased.

### B. Evaluation of the Continuous Metric of Difficulty

#### 1) Analysis of the Gameplay Data

Both the RL-based and the second rule-based methods employed the proposed difficulty metric, while the first rule-based method used the number of targets as the difficulty indicator. Users experienced a decrease in their score during a 20-trial session when the first rule-based adjustment method was adopted. Conversely, the same participants

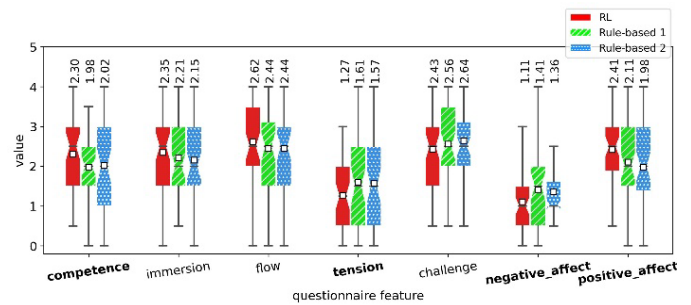

Fig. 5. Box plot players' game experience. The bold style denotes significant difference. The little squares display the mean and its value is written above the whisker. The little horizontal lines denotes the median.

---

[1] This video depicts the Visual Working Memory game and the experiment.

[2] Research Ethics Committees Certificate

Continuous Reinforcement Learning-based Dynamic Difficulty Adjustment in a Visual Working Memory Game

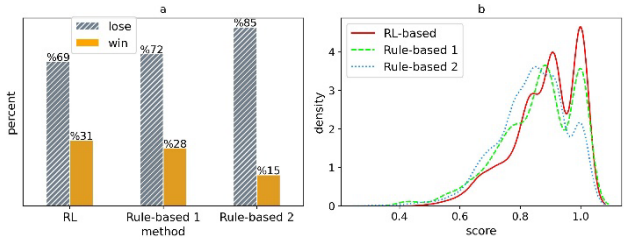

Fig. 6. (a) Percentage of wins and losses in each method. (b) Density estimation of the score.

experienced a significantly less decline in the score when the RL-based and the second rule-based methods were employed; that is, using the proposed difficulty metric for DDA led to a less decline in the score during a 20-trial session (Fig. 7). To assess this statement, correlation between scores per method and trial numbers was calculated, which represents the decline in the score. It resulted in 3 vectors, each for one method and consisting of 52 elements. T-tests between every two pairs of vectors were conducted, which showed the score decline was significantly higher in the first rule-based than in the two other methods (Table II).

TABLE I.  COMPARING GAME EXPERIENCE IN THREE DDA METHODS USING THE FRIEDMAN TEST

| *Questionnaire features* | *Statistic* | *P-value* |
|---|---|---|
| **Competence** [a] | 10.69 | **0.00** |
| Immersion | 1.93 | 0.38 |
| Flow | 2.65 | 0.27 |
| **Tension** | 10.58 | **0.01** |
| Challenge | 3.57 | 0.17 |
| **Negative affect** | 8.94 | **0.01** |
| **Positive affect** | 13.48 | **0.00** |

[a.] Bold style represents significant differences

TABLE II.  PAIRED T-TEST

| *Factor* | *Groups* | *Stat* | *P-value* | *df* |
|---|---|---|---|---|
| **Score (Avg$_{s, m}$)** | **RL vs Rule 1** | **7.21** | **0.00** | 51 |
| **Score (Avg$_{s, m}$)** | **RL vs Rule 2** | **12.73** | **0.00** | 51 |
| Score (Avg$_{s, m}$) | Rule 1 vs Rule 2 | 6.00 | 0.00 | 51 |
| **Score decline** | **RL vs Rule 1** | **3.28** | **0.00** | 51 |
| Score decline | RL vs Rule 2 | 1.03 | 0.30 | 51 |
| **Score decline** | **Rule 1 vs Rule 2** | **-2.46** | **0.01** | 51 |

*2) Analysis of the Questionnaire Data*

In order to evaluate the proposed difficulty metric in the VWM game, we only compared two rule-based methods that only differ in their indicator of task difficulty. Subjects experienced better rates of competence, immersion, tension, and negative affect in the second rule-based method compared to the first rule-based method; however, the

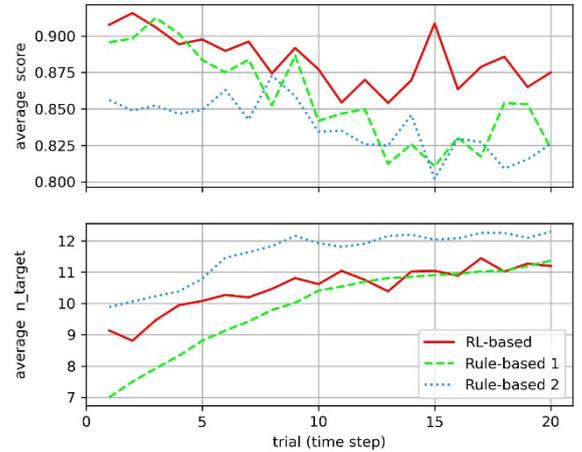

Fig. 7. Average score and challenge along trials.

Wilcoxon signed-rank test indicated that the difference was not significant.

VI. CONCLUSION

In this paper, we proposed a continuous RL-based DDA in the VWM game. Despite previous studies, we used continuous RL-based DDA in a non-competitive game with large and complex difficulty search space. We considered difficulty metric as the action of RL, and the vector of the difficulty metric and the score were considered as the state of RL. The training was performed through simulations of human players, followed by a fine-tuning phase using human subjects. A within-subject experiment involving 52 participants was conducted to compare the RL-based with two rule-based DDA methods. The proposed RL-based DDA demonstrated significant enhancements in the players' game experience, including competence, tension, and positive and negative affect. Moreover, participants achieved higher average score and win rate in the RL-based approach. Additionally, the RL-based approach kept players in higher score while the game became more challenging in a 20-trial session. The success of the RL-based DDA can be attributed to its ability to keep players at around 85% performance while making the game more challenging in a 20-trial session game. Furthermore, the incorporation of continuous RL offers several advantages in DDA. It could effectively deal with the complexity of the difficulty search space and might address the issue of biased difficulty metrics in games. However, further research is necessary to validate these hypotheses and explore additional applications of continuous RL in DDA.

Secondly, we proposed a continuous metric for the difficulty of memorization in the VWM game. Utilizing the proposed metric in DDA resulted in a significantly slighter decrease in the player's score during a 20-trial session. However, comparing two rule-based difficulty adjustment methods regarding game experience showed that the proposed continuous difficulty metric performed better, but not significantly. It was evaluated through a game experience questionnaire in a within-subject experiment involving 20 participants who played two rule-based DDA methods.

Continuous Reinforcement Learning-based Dynamic Difficulty Adjustment in a Visual Working Memory Game

For future work, the picture of the memory task can be considered as the action and the first dimension of the state. Furthermore, incorporating different dimensions of the player model into the state could potentially offer a more personalized gaming experience. Additionally, deep RL could be employed to effectively manage intricate state-action spaces, notwithstanding the greater demand for data. Finally, the proposed difficulty metric could be enhanced by adjusting the covariate in (1) using gameplay data.